\documentclass{ws-procs975x65}
\usepackage[hyperindex,plainpages=false]{hyperref}
\usepackage[depth=1,openlevel=1,numbered=true]{bookmark}

\begin{document}

\title{Finsler fluid dynamics in $\mathrm{SO}(4)$ symmetric cosmology}
\author{Manuel Hohmann$^*$}

\address{Laboratory of Theoretical Physics, Institute of Physics, University of Tartu,\\
Tartu, Tartumaa, Estonia\\
$^*$E-mail: manuel.hohmann@ut.ee\\
http://kodu.ut.ee/\textasciitilde{}manuel/}

\begin{abstract}
We discuss the most general Finsler spacetime geometry obeying the cosmological symmetry group \(\mathrm{SO}(4)\). On this background geometry we derive the equations of motion for the most general kinetic fluid obeying the same cosmological symmetry. For this purpose we propose a set of coordinates on the tangent bundle of the spacetime manifold which greatly simplifies the cosmological symmetry generators.
\end{abstract}


\bodymatter

\section{Introduction}\label{sec:intro}
We have recently derived the equations of motion for a kinetic fluid on a Finsler spacetime background, including the most general case of cosmological symmetry.\cite{Hohmann:2015ywa} This previous work provided an extension of the kinetic theory of fluids\cite{Ehlers:1971,Sarbach:2013fya,Sarbach:2013uba} to the Finsler spacetime framework.\cite{Pfeifer:2011tk,Pfeifer:2011xi,Pfeifer:2013gha} Here we discuss a particular subclass of these models, where the cosmological symmetry group is the group \(\mathrm{SO}(4)\). In this case it is possible to use a particular set of coordinates on the tangent bundle \(TM\) of the spacetime manifold \(M\) which highly simplify the derivation. The aim of this work is to introduce this new set of coordinates and to show its usefulness for describing cosmological symmetry in Finsler geometric theories of gravity.

The outline of this article is as follows. In section~\ref{sec:cosmology} we discuss the cosmological symmetry and introduce adapted coordinates on the tangent bundle \(TM\). We then use these coordinates to derive the most general Finsler geometric background obeying this symmetry in section~\ref{sec:finsler} and its observer space in section~\ref{sec:obs}. Using this background geometry, we derive the Liouville equation for a collisionless fluid in section~\ref{sec:fluid}. We end with a conclusion in section~\ref{sec:conclusion}.

\section{Cosmological Symmetry}\label{sec:cosmology}
We start with a brief review of the cosmological symmetry we intend to impose. Here we restrict ourselves to the case that the spatial geometry of the cosmological background is closed, and thus diffeomorphic to \(S^3\). We first introduce Cartesian coordinates \((X^A, A = 0, \ldots, 4)\) on the embedding space \(E = \mathbb{R}^5\), from which we derive the physical subspace
\begin{equation}
M = \{X \in E: (X^1)^2 + (X^2)^2 + (X^3)^2 + (X^4)^2 = 1\}\,.
\end{equation}
The group of cosmological symmetries on \(E\) we consider is \(\mathrm{SO}(4)\) generated by the vector fields
\begin{equation}\label{eqn:symgen}
\mathbf{R}_{\alpha} = -\epsilon_{\alpha\beta\gamma}X^{\beta}\partial_{\gamma}\,, \quad \mathbf{T}_{\alpha} = -X^4\partial_{\alpha} + X^{\alpha}\partial_4\,,
\end{equation}
where \(\alpha = 1, 2, 3\). These vector fields are tangent to \(M\), and thus leave the physical spacetime invariant. They generate the rotations and translations of \(M\), respectively.

Note that any vector field \(\xi\) on \(E\) generates a one-parameter group of diffeomorphisms \(\lambda \mapsto \varphi_{\lambda}: E \to E\). The differentials \((\varphi_{\lambda})_*: TE \to TE\) of these diffeomorphisms constitute a one-parameter group of diffeomorphisms \(\lambda \mapsto \hat{\varphi}_{\lambda} = (\varphi_{\lambda})_*\) of the tangent bundle, which is generated by a vector field \(\hat{\xi}\) on \(TE\). This vector field is called the complete lift of \(\xi\). From the coordinates \((X^A)\) on \(E\) one can construct induced coordinates
\begin{equation}
(X,Y) = Y^A\partial_A \in T_XE
\end{equation}
on \(TE\), and write the corresponding coordinate vector fields on \(TE\) as \(\partial_A =  \partial/\partial X^A, \bar{\partial}_A =  \partial/\partial Y^A\). The complete lift of a vector field \(\xi = \xi^a\partial_a\) on \(M\) then takes the form
\begin{equation}
\hat{\xi} = \xi^A\partial_A + Y^A\partial_A\xi^B\bar{\partial}_B\,.
\end{equation}
For the symmetry generating vector fields~\eqref{eqn:symgen} we thus find
\begin{equation}
\hat{\mathbf{R}}_{\alpha} = -\epsilon_{\alpha\beta\gamma}(X^{\beta}\partial_{\gamma} + Y^{\beta}\bar{\partial}_{\gamma})\,, \quad \hat{\mathbf{T}}_{\alpha} = -X^4\partial_{\alpha} + X^{\alpha}\partial_4 - Y^4\bar{\partial}_{\alpha} + Y^{\alpha}\bar{\partial}_4
\end{equation}
using induced coordinates \((X^A,Y^A)\) on \(TE\).

Since our description of fluid dynamics will be based on geometric quantities defined on the tangent bundle of \(M\), and these quantities obey cosmological symmetry in the sense that they are invariant under the action of the symmetry generating vector fields above, we introduce coordinates on \(TM\) in which the restrictions \(\hat{\mathbf{r}}_{\alpha} = \hat{\mathbf{R}}_{\alpha}|_{TM}\) and \(\hat{\mathbf{t}}_{\alpha} = \hat{\mathbf{T}}_{\alpha}|_{TM}\) of the symmetry generators simplify. Here we use coordinates \((t, y, w, \theta^+, \theta^-, \phi^+, \phi^-, \beta)\) such that
\begin{subequations}
\begin{align}
X^0 &= t\,, & (X^1, X^2, X^3, X^4)^T &= P \cdot \left(0, 0, \sin\frac{\beta}{2}, \cos\frac{\beta}{2}\right)^T\,,\\
Y^0 &= y\,, & (Y^1, Y^2, Y^3, Y^4)^T &= P \cdot w\left(0, 0, \cos\frac{\beta}{2}, -\sin\frac{\beta}{2}\right)^T\,,
\end{align}
\end{subequations}
where the matrix \(P\) is given by
\begin{multline}
P = \left(\begin{array}{cccc}
\sin\frac{\phi^-}{2} & \cos\frac{\phi^-}{2} & 0 & 0\\
-\cos\frac{\phi^-}{2} & \sin\frac{\phi^-}{2} & 0 & 0\\
0 & 0 & \sin\frac{\phi^-}{2} & \cos\frac{\phi^-}{2}\\
0 & 0 & -\cos\frac{\phi^-}{2} & \sin\frac{\phi^-}{2}
\end{array}\right) \cdot \left(\begin{array}{cccc}
\cos\frac{\phi^+}{2} & -\sin\frac{\phi^+}{2} & 0 & 0\\
\sin\frac{\phi^+}{2} & \cos\frac{\phi^+}{2} & 0 & 0\\
0 & 0 & \cos\frac{\phi^+}{2} & \sin\frac{\phi^+}{2}\\
0 & 0 & -\sin\frac{\phi^+}{2} & \cos\frac{\phi^+}{2}
\end{array}\right)\\
\cdot \left(\begin{array}{cccc}
\sin\frac{\theta^-}{2} & 0 & \cos\frac{\theta^-}{2} & 0\\
0 & \sin\frac{\theta^-}{2} & 0 & -\cos\frac{\theta^-}{2}\\
-\cos\frac{\theta^-}{2} & 0 & \sin\frac{\theta^-}{2} & 0\\
0 & \cos\frac{\theta^-}{2} & 0 & \sin\frac{\theta^-}{2}
\end{array}\right) \cdot \left(\begin{array}{cccc}
\cos\frac{\theta^+}{2} & 0 & \sin\frac{\theta^+}{2} & 0\\
0 & \cos\frac{\theta^+}{2} & 0 & \sin\frac{\theta^+}{2}\\
-\sin\frac{\theta^+}{2} & 0 & \cos\frac{\theta^+}{2} & 0\\
0 & -\sin\frac{\theta^+}{2} & 0 & \cos\frac{\theta^+}{2}
\end{array}\right)\,.
\end{multline}
In these coordinates the linear combinations \(\hat{\mathbf{j}}^{\pm}_{\alpha} = \hat{\mathbf{r}}_{\alpha} \pm \hat{\mathbf{t}}_{\alpha}\) take the simple form
\begin{subequations}\label{eqn:cosmovect}
\begin{align}
\hat{\mathbf{j}}^{\pm}_1 &= \sin\phi^{\pm}\partial_{\theta^{\pm}} + \frac{\cos\phi^{\pm}}{\tan\theta^{\pm}}\partial_{\phi^{\pm}} - \frac{\cos\phi^{\pm}}{\sin\theta^{\pm}}\partial_{\beta}\,,\\
\hat{\mathbf{j}}^{\pm}_2 &= -\cos\phi^{\pm}\partial_{\theta^{\pm}} + \frac{\sin\phi^{\pm}}{\tan\theta^{\pm}}\partial_{\phi^{\pm}} - \frac{\sin\phi^{\pm}}{\sin\theta^{\pm}}\partial_{\beta}\,,\\
\hat{\mathbf{j}}^{\pm}_3 &= -\partial_{\phi^{\pm}}\,.
\end{align}
\end{subequations}
These are the coordinates we will be working with in the following sections.

\section{Finsler Geometric Background}\label{sec:finsler}
We now discuss the background geometry we need for our description of fluid dynamics. Note that the tangent bundle \(TM\) itself already comes with a number of geometric structures, which are useful for defining Finsler geometry without using induced coordinates.\cite{Bucataru} Most important for our construction are the tangent and cotangent structures, which can be defined as follows. Let \(\pi: TM \to M\) be the bundle map of \(TM\). Its differential \(\pi_*\) assigns to every vector \(v \in T_pTM\) a vector \(\pi_*(v) \in T_{\pi(p)}M\). Since both \(p\) and \(\pi_*(v)\) are elements of the same tangent vector space \(T_{\pi(p)}M\), we can construct a curve \(\lambda \mapsto p + \lambda\pi_*(v)\) on \(TM\) and define
\begin{equation}
J(v) = \left.\frac{d}{d\lambda}(p + \lambda\pi_*(v))\right|_{\lambda = 0} \in T_pTM
\end{equation}
as the tangent vector of this curve at \(\lambda = 0\). The map \(J: TTM \to TTM\) is called the tangent structure. It can be represented as a $(1,1)$-tensor field on \(TM\), which can be written in matrix form as
\begin{equation}
J = \frac{1}{2w}\left(\begin{array}{cccccccc}
0 & 0 & 0 & 0 & 0 & 0 & 0 & 0\\
2w & 0 & 0 & 0 & 0 & 0 & 0 & 0\\
0 & 0 & 0 & 0 & 0 & w\cos\theta^+ & w\cos\theta^- & w\\
0 & 0 & 0 & 0 & -\sin\beta & -\sin\theta^+ & \cos\beta\sin\theta^- & 0\\
0 & 0 & 0 & -\sin\beta & 0 & \cos\beta\sin\theta^+ & -\sin\theta^- & 0\\
0 & 0 & 0 & \frac{1}{\sin\theta^+} & \frac{\cos\beta}{\sin\theta^+} & 0 & \frac{\sin\beta\sin\theta^-}{\sin\theta^+} & 0\\
0 & 0 & 0 & \frac{\cos\beta}{\sin\theta^-} & \frac{1}{\sin\theta^-} & \frac{\sin\beta\sin\theta^+}{\sin\theta^-} & 0 & 0\\
0 & 0 & 0 & -\frac{\cos\beta}{\tan\theta^-} - \frac{1}{\tan\theta^+} & -\frac{\cos\beta}{\tan\theta^+} - \frac{1}{\tan\theta^-} & -\frac{\sin\beta\sin\theta^+}{\tan\theta^-} & -\frac{\sin\beta\sin\theta^-}{\tan\theta^+} & 0
\end{array}\right)
\end{equation}
in the coordinate basis on \(TM\). Its dual map \(J^*: T^*TM \to T^*TM\) defined by
\begin{equation}
J^*(\sigma)(v) = \sigma(J(v))
\end{equation}
is called the cotangent structure. Its matrix representation is given by the transpose of the matrix representation of \(J\).

Another structure can be derived from the fact that every element \(p \in TM\), being an element of a vector space \(T_xM\) for some \(x \in M\), can be multiplied by a real number. This induces a one-parameter group of diffeomorphisms \(\chi_{\lambda}: TM \to TM\) as \(\chi_{\lambda}(p) = e^{\lambda}p\). The generating vector field of \(\chi_{\lambda}\) is called the Liouville vector field \(\mathbf{c}\). It takes the form
\begin{equation}
\mathbf{c} = y\partial_y + w\partial_w
\end{equation}
in the coordinates we are using.

We now equip \(TM\) with the most general Finsler geometry which obeys cosmological symmetry as defined in the previous section. Our starting point is the Finsler function \(F: TM \to \mathbb{R}^+\), which is one-homogeneous, continuous on \(TM\) and smooth on the space \(\widetilde{TM}\) defined as \(TM\) without the null structure.\cite{Pfeifer:2011tk,Pfeifer:2011xi,Pfeifer:2013gha} Using the vector fields introduced above, the conditions of homogeneity and cosmological symmetry read
\begin{equation}
\mathcal{L}_{\mathbf{c}}F = F\,, \quad \mathcal{L}_{\hat{\mathbf{j}}^{\pm}_{\alpha}}F = 0\,.
\end{equation}
The most general Finsler function satisfying these conditions takes the form
\begin{equation}\label{eqn:finsler}
F = F(t, y, w) = y\tilde{F}(t, w/y)
\end{equation}
for some smooth function \(\tilde{F}\). We then construct the Cartan one-form
\begin{equation}
\theta = \frac{1}{2}J^*\left(dF^2\right) = \tilde{F}\left[(y\tilde{F} - w\tilde{F}_w)dt + \frac{y}{2}\tilde{F}_w\left(\cos\theta^+d\phi^+ + \cos\theta^-d\phi^- + d\beta\right)\right]\,,
\end{equation}
where the subscript \(w\) indicates a derivative with respect to the second argument of \(\tilde{F}\). Its exterior derivative \(\omega = d\theta\), called the Cartan two-form, is a symplectic form on \(\widetilde{TM}\). It can be used to define the geodetic spray as the unique vector field \(\mathbf{s}\) on \(\widetilde{TM}\) such that
\begin{equation}
\iota_{\mathbf{s}}\omega = -\frac{1}{2}dF^2\,.
\end{equation}
This finally yields
\begin{equation}\label{eqn:spray}
\mathbf{s} = y\partial_t + 2w\partial_{\beta} - y^2\frac{\tilde{F}_t\tilde{F}_{ww} - \tilde{F}_w\tilde{F}_{tw}}{\tilde{F}\tilde{F}_{ww}}\partial_y - \frac{y^2\tilde{F}\tilde{F}_{tw} + yw\tilde{F}_t\tilde{F}_{ww} - yw\tilde{F}_w\tilde{F}_{tw}}{\tilde{F}\tilde{F}_{ww}}\partial_w\,.
\end{equation}
The geodetic spray has a simple geometric interpretation. Its integral curves are the canonical lifts of curves on \(M\) which minimize the Finsler length functional and correspond to the trajectories of freely falling test masses.

\section{Observer Space}\label{sec:obs}
In the previous section we have discussed the Finsler background geometry on (almost) the whole tangent bundle \(TM\). For the kinetic theory of fluids, however, we are interested only in those tangent vectors which correspond to physically allowed four-velocities. These constitute a submanifold \(O \subset TM\), which we call observer space. Its elements are future timelike and normalized by the Finsler function~\eqref{eqn:finsler}, such that \(F|_O = 1\). Further, the geodetic spray \(\mathbf{s}\) is tangent to \(O\). Its restriction \(\mathbf{r} = \mathbf{s}|_O\) is called the Reeb vector field.

In order to derive the Reeb vector field, which is necessary for the description of fluid dynamics, we first introduce new coordinates \((\tilde{t}, \tilde{y}, \tilde{w}, \tilde{\theta}^+, \tilde{\theta}^-, \tilde{\phi}^+, \tilde{\phi}^-, \tilde{\beta})\) on \(TM\) such that
\begin{gather}
\tilde{t} = t\,, \quad \tilde{y} = y\tilde{F}\left(t,\frac{w}{y}\right)\,, \quad \tilde{w} = \frac{w}{y}\,,\nonumber\\
\tilde{\theta}^+ = \theta^+\,, \quad \tilde{\theta}^- = \theta^-\,, \quad \tilde{\phi}^+ = \phi^+\,, \quad \tilde{\phi}^- = \phi^-\,, \quad \tilde{\beta} = \beta\,.
\end{gather}
In these coordinates the geodetic spray~\eqref{eqn:spray} takes the simple form
\begin{equation}
\mathbf{s} = \frac{\tilde{y}}{\tilde{F}}\left(\tilde{\partial}_t + 2\tilde{w}\tilde{\partial}_{\beta} - \frac{\tilde{F}_{tw}}{\tilde{F}_{ww}}\tilde{\partial}_w\right)\,.
\end{equation}
These coordinates further have the advantage that the observer space \(O\) is given as a connected component of the submanifold \(\tilde{y} = 1\), so that one can use the remaining seven coordinates to parametrize \(O\). The Reeb vector field, being the restriction of the geodetic spray to this submanifold, thus reads
\begin{equation}\label{eqn:reeb}
\mathbf{r} = \frac{1}{\tilde{F}}\left(\tilde{\partial}_t + 2\tilde{w}\tilde{\partial}_{\beta} - \frac{\tilde{F}_{tw}}{\tilde{F}_{ww}}\tilde{\partial}_w\right)\,.
\end{equation}
This will be the central ingredient for our definition of fluid dynamics in the next section.

\section{Fluid Dynamics}\label{sec:fluid}
We finally come to the discussion of fluid dynamics on the Finsler spacetime background derived above, making use of the kinetic theory of fluids.\cite{Ehlers:1971,Sarbach:2013fya,Sarbach:2013uba} Since the background geometry obeys cosmological symmetry, the canonical lifts~\eqref{eqn:cosmovect} are tangent to the observer space \(O\). A fluid obeys the same symmetry if and only if its one-particle distribution function \(\phi\) is invariant under the restriction of these canonical lifts to \(O\). In the present case the most general one-particle distribution function satisfying this condition takes the form \(\phi = \phi(\tilde{t},\tilde{w})\). Its Lie derivative with respect to the Reeb vector field~\eqref{eqn:reeb} is thus given by
\begin{equation}
\mathcal{L}_{\mathbf{r}}\phi = \frac{1}{\tilde{F}}\left(\phi_t - \frac{\tilde{F}_{tw}}{\tilde{F}_{ww}}\phi_w\right)\,.
\end{equation}
For the simplest possible case of a collisionless fluid the equations of motion hence take the form
\begin{equation}
\tilde{F}_{ww}\phi_t = \tilde{F}_{tw}\phi_w\,.
\end{equation}
This is the Liouville equation for a fluid with cosmological symmetry.

\section{Conclusion}\label{sec:conclusion}
We have derived the equations of motion for a collisionless fluid on a Finsler background geometry obeying the cosmological symmetry group \(\mathrm{SO}(4)\). For this purpose we introduced coordinates on the tangent bundle of the spacetime manifold, in which the generators of the cosmological symmetry simplify significantly, and expressed all relevant Finsler geometric quantities in these coordinates.

The result we present here is the first step towards understanding the cosmological dynamics of Finsler geometric theories of gravity. The basic idea behind this research is to consider a kinetic fluid, as used in this article, as the source of gravity, in analogy to the classical perfect fluid in metric cosmology. A necessary ingredient, besides the fluid dynamics presented here, is a thorough understanding of the most general Finsler geometric spacetime with cosmological symmetry. By introducing a suitable set of coordinates, we provide a way to achieve this understanding.

Using a different choice of coordinates, this work can easily be generalized. This will allow a description of fluid dynamics on Finsler geometric backgrounds whose cosmological symmetry is given by the groups \(\mathrm{SO}(3,1)\) and \(\mathrm{ISO}(3)\), corresponding to hyperbolic and flat spatial geometries. These will be discussed in future work.

\section*{Acknowledgments}
The author is happy to thank Christian Pfeifer for extensive discussions and fruitful collaboration. He gratefully acknowledges the full financial support of the Estonian Research Council through the Postdoctoral Research Grant ERMOS115 and the Startup Research Grant PUT790.

\end{document}